\documentclass[journal]{IEEEtran}
\usepackage{amsmath,amsfonts}
\usepackage{algorithmic}
\usepackage{algorithm}
\usepackage{array}
\usepackage{tikz}
\usetikzlibrary{shapes.geometric, intersections, positioning,arrows}
\usepackage[caption=false,font=normalsize,labelfont=sf,textfont=sf]{subfig}
\usepackage{textcomp}
\usepackage{stfloats}
\usepackage{url}
\usepackage{verbatim}
\usepackage{graphicx}
\usepackage{hyperref} 
\usepackage{cite}

\usepackage{listings}
\usepackage{xcolor}

\usepackage{adjustbox} 

\usepackage{pgfplots}
\pgfplotsset{compat=1.17}

\begin{document}

\title{DCMF: A Dynamic Context Monitoring and Caching Framework for Context Management Platforms}

\author{%
    Ashish Manchanda,
    Prem Prakash Jayaraman,
    Abhik Banerjee,
    Kaneez Fizza,
    and Arkady Zaslavsky%
    \thanks{Corresponding author: Ashish Manchanda (e-mail: amanchanda@swin.edu.au).}
    \thanks{Ashish Manchanda, Prem Prakash Jayaraman, Abhik Banerjee, and Kaneez Fizza are with School of Science, Computing, Engineering and Technology, Swinburne
University of Technology, Melbourne, VIC 3122, Australia (e-mails: amanchanda@swin.edu.au; pjayaraman@swin.edu.au; abanerjee@swin.edu.au; kfizza@swin.edu.au).}%
    \thanks{Arkady Zaslavsky is with the School of Information Technology, Deakin University, Melbourne, VIC, Australia (e-mail: arkady.zaslavsky@deakin.edu.au).}%

}

\markboth
{DCMF: A Dynamic Context Monitoring and
Caching Framework for Context Management
Platforms}
{DCMF: A Dynamic Context Monitoring and
Caching Framework for Context Management
Platforms}


\maketitle

\begin{abstract}

The growth of context-aware Internet of Things (IoT) applications has strengthened the need for prompt and accurate context information. Context is derived by aggregating and inferring from dynamic IoT data, making it even more dynamic \& volatile, posing significant challenges for maintaining freshness and real-time accessibility. One effective solution to address these challenges is caching, however, the dynamic and transient nature of context in IoT environments introduce complexities that traditional caching policies cannot handle effectively(e.g., ensuring real-time accessibility for high-frequency queries, maintaining freshness of rapidly changing data, or handling critical updates). 
To address these challenges, we propose a Dynamic Context Monitoring Framework (DCMF), designed to enhance context caching within Context Management Platforms (CMPs) by dynamically monitoring and managing context. DCMF consists of two integral components: the Context Evaluation Engine (CEE) and the Context Management Module (CMM). The CEE evaluates the Probability of Access (PoA) using parameters such as Quality of Service (QoS), Quality of Context (QoC), Cost of Context (CoC), timeliness, and Service Level Agreements (SLAs), assigning weights that reflect relevance and access likelihood. Based on CEE’s output, the CMM uses a hybrid Dempster-Shafer approach to manage Context Freshness (CF), adjusting belief levels and confidence scores to decide whether to cache, evict, or refresh context. We implemented DCMF in Context as a Service (CoaaS) platform and evaluated it with real-world data from smart city applications, specifically focusing on traffic management and roadwork scenarios. The results show that DCMF achieves a 12.5\% higher cache hit rate compared to state of the art techniques like Context Aware Caching (m-CAC) which is based on context popularity and relevance to maximize cache utility. Additionally, DCMF reduces the cache expiry ratio by up to 60\% when compared to m-CAC, ensuring timely delivery of relevant context and minimizing latency. These improvements highlight DCMF's scalability and effectiveness in supporting dynamic context-aware IoT applications.

\end{abstract}

\begin{IEEEkeywords}
context freshness, probability of access, context as a service, Internet of Things, dynamic hybrid approach
\end{IEEEkeywords}

\section{Introduction}

\IEEEPARstart{I}{oT} applications plays vital role in collecting and processing data from various IoT sensors to provide insights that support real-time decision-making and automation processes. Data from IoT sensors is critical for inferring context, which enables IoT applications to offer relevant and personalised services to users. Context, as defined by Anind Dey~\cite{dey2001understanding}, is ``any information that can be used to characterize the situation of an entity. An entity is a person, place, or object that is considered relevant to the interaction between a user and an application, including the user and applications themselves''.

While the need for real-time context information grows, the dynamic and transient nature of IoT data poses significant challenges. Context change rapidly, requiring efficient management to maintain accuracy and relevance. Delays or inefficiencies in accessing context can lead to degraded user experiences and system failures. Caching is an effective solution to mitigate these issues, ensuring prompt delivery of context. However, traditional caching strategies, designed for static or predictable data, struggle in dynamic IoT environments due to their reliance on metrics like frequency or size~\cite{s23218779, Al-Ward2022Caching}. This necessitates novel strategies that adapt to the evolving demands of IoT applications.

To address these challenges of efficient context sharing ecosystem for IoT and management, Context Management Platforms (CMPs) such as CoaaS, Nexus, and FIWARE Orion have been proposed in the literature as middleware solutions~\cite{perera2014context,8480240,fiwareorion2022,lehmann2004from}. These platforms acts as aggregators \& facilitate the exchange of context between Context Providers (CPs) any IoT device or system that can provide context and Context Consumers (CCs) IoT devices or applications that require context for their functioning. CMPs process context queries from consumers and coordinate with multiple providers to deliver the required context information efficiently.

\begin{figure}
    \centering
    \includegraphics[width=9cm]{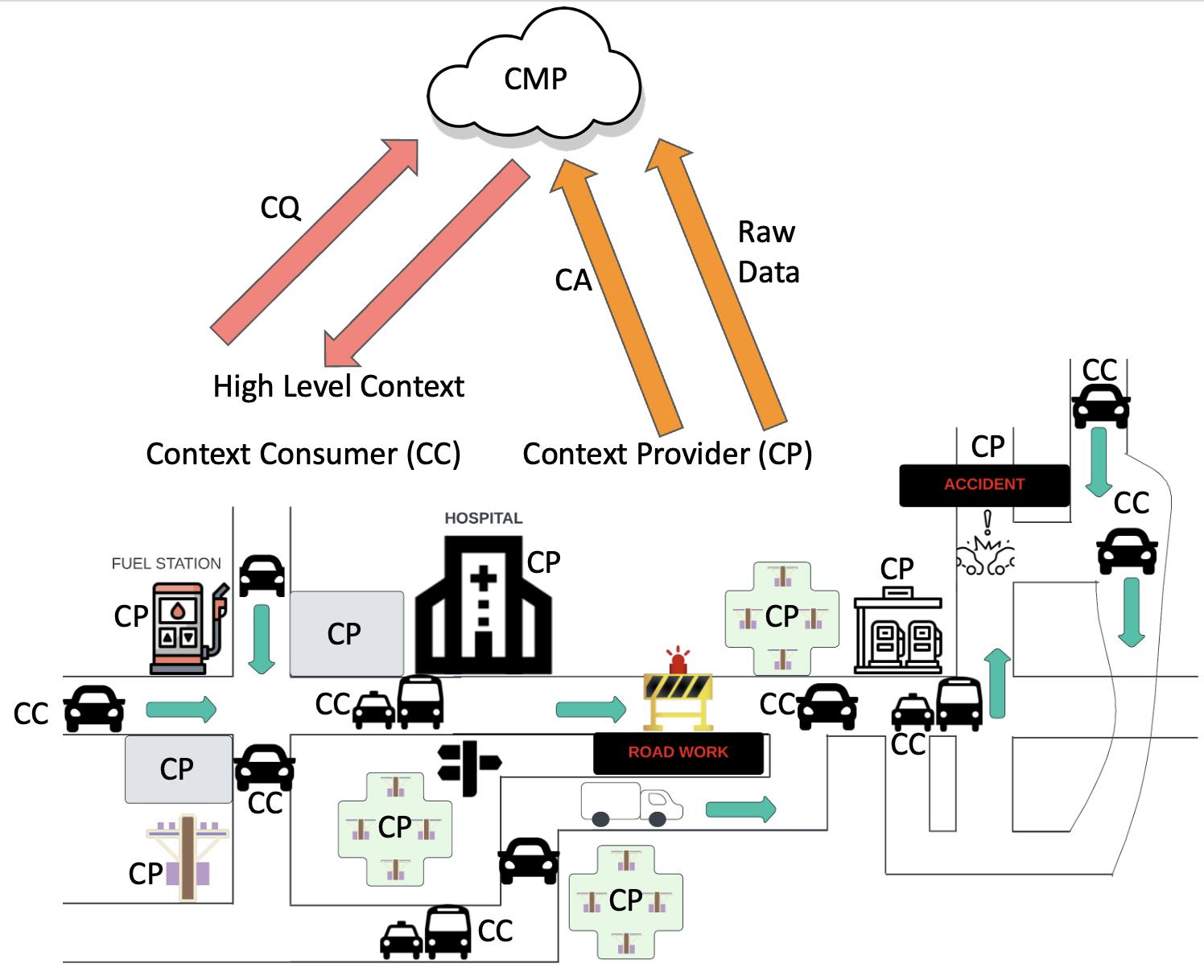}
    \caption{Context Sharing Ecosystem for Road Work/Hazard Scenario in Smart City.}
    \label{fig:ecosystem}
\end{figure}

In Figure~\ref{fig:ecosystem}, CCs send queries called \textit{Context Queries} (CQs) that may consist of multiple pieces of information. CMPs leverage caching mechanisms to reduce latency and improve responsiveness. However, the effectiveness of this caching mechanism depends heavily on the dynamic nature of IoT contexts. When cache miss occur, the system must retrieve context from CPs, leading to increased latency. Our research shows that traditional caching strategies struggle to maintain optimal performance in highly dynamic IoT environments, where context values can change rapidly and unpredictably~\cite{perera2014context}. This challenge necessitates a more sophisticated approach to context caching that can adapt to the volatile nature of IoT applications.

\subsection{Motivating Scenario}~\label{sec:motivating}

Traffic management in cities is a significant challenge, particularly when unexpected events, such as roadworks, occur. Consider a scenario of a context-aware IoT navigation application in autonomous vehicles (AVs) in a smart city, as shown in Figure~\ref{fig:ecosystem}. In this scenario AVs are entities that queries CMP for real time updates. The context includes road conditions, obstacles like roadwork, and traffic congestion. The application uses data from vehicle sensors, as well as information from other vehicles and IoT applications. For example, if an IoT application detects a road sign or obstacle, it sends this information to a CMP. This information is then aggregated into context items and shared with all connected vehicles in real-time. Consequently, another car approaching the same area knows about the obstacle in advance, enabling safer and more efficient navigation.

When such context-aware IoT applications embedded in vehicles query the CMP for road conditions, the lack of an effective caching mechanism leads to significant inefficiencies. The CMP must repeatedly request context information from CPs for each query, even when multiple applications could reuse the same context stemming from IoT data. This process not only introduces delays due to the time required to fetch data from distributed sources but also increases the communication overhead for the CMP and the CPs. For instance, querying a traffic database or weather station for identical information multiple times results in unnecessary latency, reduced responsiveness, and increased network load. Moreover, as the fetched data is often dynamic, such as road conditions or traffic congestion, its relevance may diminish during the delay, leading to applications receiving outdated or less accurate context information. These challenges highlight the critical need for a dynamic caching strategy that can store and reuse interpreted context, thereby minimizing repeated queries, reducing overheads, and ensuring real-time and accurate responses, thus laying the groundwork for our research.





Traditional caching strategies, often designed for static data, struggle with the complexities of dynamic IoT environments, leading to suboptimal performance~\cite{Al-Ward2022Caching}. Unlike static data, IoT data changes rapidly, and the inferred context, requires continuous updates to ensure it remains accurate and relevant. The challenge lies in determining which context to cache, when to refresh or evict it, and how to ensure the cache contains only the most useful context for CCs. This necessitates a dynamic caching framework that can handle the fast-changing demands of IoT applications.


To address these challenges, we propose, implement, and evaluate the Dynamic Context Monitoring Framework (DCMF), which enhances caching efficiency for CMPs through two key components.

\begin{enumerate} 

    \item  Prioritized Context Caching for Real-Time Applications: 
    Introduces a framework that identifies and assigns weights to CAs, enabling IoT applications to make intelligent decisions about which CIs to cache, refresh, or evict. This prioritization ensures low latency, high responsiveness, and effective resource utilization.

    \item Probabilistic and Utility-Driven Context Management: Incorporates MAUT and probabilistic approaches, such as the DST, to dynamically adjust caching decisions based on both real-time and historical access patterns. By utilizing dynamic metrics such as CF \& PoA to balance the trade-offs between maintaining relevance and minimizing unnecessary updates, reducing cache expiry ratio by 60\% compared to state-of-the-art methods.

    \item Comprehensive Evaluation on Real-World Data: Validates the proposed framework through extensive experiments using real-world roadwork data, demonstrating improvements in cache hit ratio, reduced latency, and efficient handling of over 70,000 daily context requests.

    \item Scalable and Adaptive Solution for IoT Context Management: Demonstrates scalability by maintaining optimal performance under varying workloads and high query rates, meeting the stringent response time requirements of real-time IoT applications.

\end{enumerate}

The remainder of this paper is organized as follows. Section \ref{sec:literature} reviews related work. Section~\ref{sec:Design} describes the design and implementation of the DCMF. Section~\ref{sec:evaluation} presents an evaluation of the framework, and Section~\ref{sec:result} presents the results and concludes the study.




\section{Related Work}~\label{sec:literature}

Research in the field of IoT has emphasized the critical role of data caching in improving system performance, particularly within the domain of context-aware computing. Traditional caching strategies, such as Least Recently Used (LRU) and Least Frequently Used (LFU), have been widely employed in computing environments~\cite{weerasinghe2023traditional,8362880}. However, their effectiveness deteriorates in dynamic IoT scenarios due to the inferred, aggregated, and rapidly changing nature of context~\cite{rizwan2016realtime,bettini2010context}. These conventional methods often neglect real-time variations in the relevance and timeliness of data, resulting in inefficiencies in retrieval processes and suboptimal resource utilization~\cite{manchanda2023hybrid,Li2015,ETSI2018,fiwareorion2022}.

Several studies have proposed adaptive caching strategies to address these challenges for context-aware IoT applications. For example, Khargharia et al.~\cite{khargharia2022probabilistic} introduced ConCaf, which leverages context query logs and machine learning models to probabilistically estimate the demand for context information. While ConCaf improves the response time of CMPs by preemptively caching context based on anticipated queries, its primary focus is on query demand estimation. This limits the system’s ability to dynamically prioritize context attributes, which is critical in volatile IoT environments.

Proactive caching approaches have also been explored to enhance performance through situation prediction. For instance, the study in~\cite{s23104767} uses historical data to predict future context needs, reducing cache misses and improving response times. However, this approach lacks dynamic attribute prioritization and adaptive cache management, making it less effective in rapidly evolving IoT scenarios where new context attributes may emerge unpredictably.

Despite these advancements, the dynamic and multi-dimensional nature of context stemming from IoT data presents unresolved challenges. Existing caching strategies often overlook the need for real-time monitoring and adaptation, which are essential to accommodate the fluctuating demands of IoT environments~\cite{weerasinghe2023adaptive,weerasinghe2023context}. Moreover, the hierarchical and interdependent nature of context attributes in CMPs further complicates caching decisions, as updates to a single attribute can affect multiple dependent contexts~\cite{MEDVEDEV2020}.


Our research addresses the limitations of existing caching strategies in IoT by introducing a holistic framework that integrates real-time monitoring and adaptive prioritization based on CF and PoA. While prior methods such as m-CAC, m-Greedy, and m-Myopic have been used as benchmarks, each focuses on a single caching parameter—popularity, frequency, or recency, respectively—and lacks adaptability to the multi-dimensional and dynamic nature of context in IoT systems~\cite{Muller2017}. In contrast, our proposed DCMF framework leverages MAUT and DST to enable decision-making, ensuring that only the most relevant and timely context is cached. This results in reduced latency, minimized expiry, and improved responsiveness, establishing DCMF as a comprehensive and state-of-the-art approach for context caching.

\section{Dynamic Context Management Framework (DCMF)} \label{sec:Design}

\begin{figure} 
\centering
\includegraphics[width=9cm]{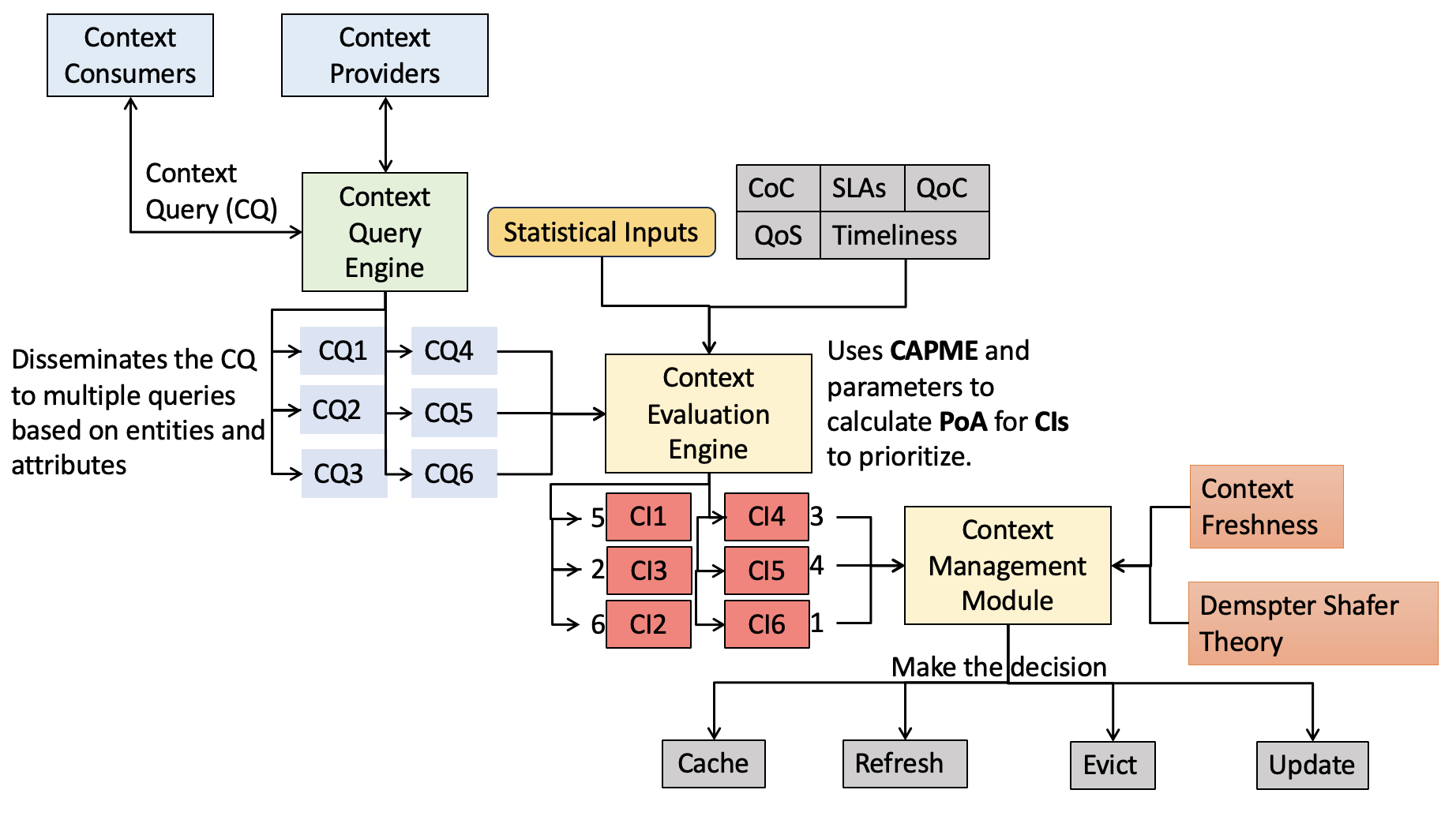}
\caption{Overview \& Architecture of Dynamic Context Monitoring Framework (DCMF).
} \label{DCMF}
\end{figure}

In this section, we present the design and methodology of the proposed DCMF, which aims to enhance caching efficiency in CMPs for IoT applications. The DCMF comprises two key components as shown in Figure~\ref{DCMF}.

\begin{itemize}
    \item Context Evaluation Engine (CEE): Computes and prioritizes context items (CIs) and their associated context attributes (CAs) to optimize caching decisions.
    \item Context Management Module (CMM): Leverages Dempster-Shafer Theory (DST) to continuously evaluate Context Freshness (CF) and Probability of Access (PoA) at runtime, enabling dynamic and efficient decisions on updating, evicting, or refreshing context.
\end{itemize}

We adopt Context Spaces Theory (CST)~\cite{1276902} for its ability to model context as multidimensional spaces, enabling efficient representation and reasoning about dynamic and complex contexts. CST supports both numerical and semantic attributes, providing flexibility for diverse IoT scenarios. By leveraging situation subspaces and their operations, CST enhances decision-making in our framework, ensuring adaptive and efficient context management. CST defines context at multiple levels:
\begin{itemize}
    \item Context Attribute (CA): Represents the lowest-level information, such as IoT data or sensor readings.
    \item Context Item (CI): Aggregates multiple context attributes to provide a higher-level abstraction, reflecting the environment's state. 
\end{itemize}

A critical aspect of such a framework is the continuous monitoring of key metrics, particularly CF~\cite{manchanda2023hybrid} \& PoA~\cite{10643912}. These metrics determine whether context within the CMP should be cached anew, evicted, or maintained in the cache memory. The dynamic nature of context and the importance of maintaining its freshness require specialized strategies designed specifically for IoT applications~\cite{al2015internet,ryan2000from}.

Each CI is inherently dynamic, reflecting real-time changes in IoT environments, and may serve as an attribute for higher-level contexts, enabling adaptive and complex inferences in rapidly evolving conditions.

In dynamic IoT environments, CAs represent fundamental pieces of information, such as sensor data or external metadata, which collectively define a higher-level CI. Each \( CI_i \) is formed by aggregating multiple relevant \( CA \) values, which can be expressed mathematically as in Eq~\ref{eq:context_item}.
\begin{equation}
    CI_i = f(CA_1, CA_2, \dots, CA_N)
    \label{eq:context_item}
\end{equation}

Here, \( f \) denotes a function or rule set that specifies how the CAs \( CA_1 \) to \( CA_N \) are combined to infer \( CI_i \). This abstraction enables the modeling of real-world scenarios, where \( CAs \) may dynamically change over time due to fluctuating environmental conditions, sensor inputs, or user actions.

As discussed in Section~\ref{sec:motivating}, the roadwork scenario includes the following key context attributes \( CA_1 \): Presence of Road Work Sign (Boolean), \( CA_2 \): Speed Limit (Numeric, e.g., km/h), \( CA_3 \): Location (Coordinates, e.g., latitude and longitude), \( CA_4 \): Congestion Level (Numeric, e.g., vehicles per km), \( CA_5 \): Time of Day (Timestamp).

The roadwork context \( CI_{\text{roadwork}} \) can dynamically change over time and may include attributes like:

\begin{equation}
\begin{split}
    CI_{\text{roadwork}} = (&\text{traffic\_density}, \text{detour\_routes}, \\
    &\text{construction\_duration}, \text{proximity\_to\_roadwork}, \\
    &\text{road\_closure\_duration}, \text{weather\_conditions})
\end{split}
\label{eq:roadwork_context}
\end{equation}

An example of inferring the roadwork \(CI\) is as follows:

\begin{equation}
\begin{aligned}
    &\text{If } CA_1 = \text{True} \text{ AND } CA_2 \leq \text{Reduced Speed Limit} \\
    &\text{AND } CA_4 \geq \text{High Congestion Level}, \\
    &\text{Then } CI_{\text{roadwork}} = \text{True}.
\end{aligned}
\label{eq:infer1}
\end{equation}

Here, the \( CI_{\text{roadwork}} \) evolves dynamically, reflecting changes in the relevant attributes. CIs can also serve as attributes for higher-level contexts. For example, a \textit{Hazard} \( CI_{\text{hazard}} \) include; Roadwork (\( CI_{\text{roadwork}} \)), \( CA_6 \): Raining (Boolean), \( CA_7 \): Visibility Level (Numeric, e.g., meters).

\begin{equation}
    CI_{\text{hazard}} = g(CI_{\text{roadwork}}, CA_6, CA_7)
    \label{eq:hazard_context}
\end{equation}

The hazard context is defined in Equation~\ref{eq:hazard_context}, where \( g \) defines how the roadwork context and additional attributes combine to infer the hazard context. The evolving nature of CAs ensures that the PoA and CF of the inferred contexts are maintained. This is critical for the DCMF to make timely and informed decisions about caching, eviction, or refresh operations as shown in Equation~\ref{eq:Infer}.

\begin{equation}
\begin{aligned}
    &\text{If } CI_{\text{roadwork}} = \text{True} \text{ AND } CA_6 = \text{True} \\
    &\text{AND } CA_7 \leq \text{Low Visibility Threshold}, \\
    &\text{Then } CI_{\text{hazard}} = \text{High Hazard}.
\end{aligned}
\label{eq:Infer}
\end{equation}



Each \( CA \) has an acceptable range \( R_{CA_1} \): Roadwork sign present (\( CA_1 = \text{True} \)), \( R_{CA_2} \): Reduced speed limit (\( CA_2 \leq \text{Reduced Speed Limit} \)),  \( R_{CA_4} \): High congestion (\( CA_4 \geq \text{High Congestion} \)). The intersection of these regions defines \( CI_{\text{RW}} \) as:

\begin{equation}
\begin{split}
CI_{RW} = \Big\{ \mathbf{CA}_{CI_{RW}} \ \Big| \ CA_1 \in R_{CA_1}, \\
CA_2 \in R_{CA_2}, \ CA_4 \in R_{CA_4} \Big\}
\end{split}
\label{eq:ci_rw}
\end{equation}

\subsection{Context Evaluation Engine (CEE)}

The CEE is responsible for computing the utility of each \( CI \) based on its parameters and determining its PoA\cite{10643912}. This ensures that the most relevant CIs are sent to the CMM for actions decision making. The CEE uses MAUT~\cite{Dyer2016} to evaluate and prioritize CIs, optimizing cache space allocation.


The utility score of a \( CI_i \) is computed as \( U(CI_i) = \sum_{j=1}^{m} w_j \cdot u_j(CI_i) \)\label{eq:utility}, where \(w_j\) is the weight of the \(j\)-th attribute, indicating its relative importance, \(u_j(C_i)\) is the utility score of \(CI_i\) based on the \(j\)-th attribute and \(\sum_{j=1}^{m} w_j = 1\) ensures that the weights are normalized, balancing the influence of each attribute. The weights \( w_j \) are determined through AHP~\cite{10734123}.

PoA is another key metric that estimates the likelihood of a CI being accessed based on historical access patterns and real-time CQs. This helps the CMP make informed decisions about which CIs to cache, evict, or refresh. PoA for a \(CI_i\) is computed as:

\begin{equation} 
\text{PoA}(CI_i) = \alpha \cdot \frac{h(CI_i)}{\sum_{k=1}^{n} h(CI_k)} + (1 - \alpha) \cdot \frac{q(CI_i)}{N} 
\label{eq:poa} 
\end{equation}

where \(h(CI_i)\) is the historical frequency of accesses to \(CI_i\), \(q(CI_i)\) is the number of recent CQs for \(CI_i\) within a time window \(T\), \(N\) is the total number of CQs in \(T\) and \( \alpha \in [0, 1] \) (where \( \alpha \in \mathbb{R} \) and bounded) is a balancing factor between historical and real-time trends. The parameter \( \alpha \) varies between 0 and 1 inclusive. When \( \alpha \to 1 \), the model emphasizes historical access patterns; conversely, when \( \alpha \to 0 \), real-time trends dominate. The statistical inputs \( h(CI_i) \) and \( q(CI_i) \) are modeled using Poisson and Gaussian distributions, respectively, as these provide a straightforward representation of historical access rates and recent query trends, facilitating easier computation and integration into the priority scoring framework.

The PoA for a \(CI_i\) is calculated as the ratio of its total queries to the overall queries, incorporating both direct requests for \(CI_i\) and inferred CAs. This ensures that both direct and indirect contributions to access patterns are considered in the caching decision.

The Algorithm~\ref{alg:cee} reflects how the CEE optimizes caching by evaluating both the utility and PoA of CIs. It prioritizes items based on their combined score and ensures that the cache contains the most relevant and frequently accessed CI. By dynamically adjusting to changing access patterns, the CEE ensures efficient caching, leading to improved system performance and reduced latency.

\begin{algorithm}[H]
\caption{Context Evaluation and Prioritization in the CEE}
\label{alg:cee}
\begin{algorithmic}[1]
\REQUIRE Set of context items $CI = \{CI_1, CI_2, \ldots, CI_n\}$, Attributes $CA = \{CA_1, CA_2, \ldots, CA_m\}$, Historical data $H$, Recent queries $Q$
\ENSURE Prioritized list of context items
\STATE Initialize $PriorityList \gets \emptyset$
\FOR{each $C_i \in CI$}
    \STATE Compute utility $U(C_i)$ using Eq.~\ref{eq:utility}
    \STATE Compute PoA $PoA(C_i)$ using Eq.~\ref{eq:poa}
    \STATE $Priority(C_i) \gets \beta \cdot U(C_i) + (1 - \beta) \cdot PoA(C_i)$
\ENDFOR
\STATE Sort context items $CI$ by $Priority(C_i)$ in descending order
\STATE Populate $PriorityList$ with sorted context items
\STATE \textbf{Return} $PriorityList$
\end{algorithmic}
\end{algorithm}

\subsection{Context Management Module (CMM)}

The CMM is responsible for efficiently managing cached context by ensuring that CIs or CAs are both relevant and fresh. The CMM combines the PoA and CF metrics using DST to dynamically decide whether to cache, update, evict, or refresh CIs \& CAs. The synergy of PoA and CF ensures that the cache remains both high-performing and up-to-date, adapting to fluctuating access patterns based on spatial and the temporal nature of IoT data. 

CF monitors the age of the context. It indicates whether a CI or CA is still valid or has become stale. 

\begin{equation}
    \text{CF}(CI_i) = e^{-\lambda \cdot \Delta t_i}
    \label{eq:cf}
\end{equation}

where \( \Delta t_i \) is the time since the last update of \( CI_i \) and \( \lambda \) is a decay constant, representing how quickly the context becomes stale.

A CF close to 1 means the context is fresh, while a CF close to 0 suggests that the context has aged and may require refreshing or based on PoA it needs to be evicted to save the cost and memory. The DST offers a mechanism to combine evidence from both PoA and CF to compute a belief score for context. This belief score reflects the likelihood that a context should remain cached, be refreshed, or be evicted.

The combined belief \( m_{\text{Combined}}(CI_i) \) for a CI \( CI_i \) is given by:

\begin{equation}
    m_{\text{Combined}}(CI_i) = \frac{1}{1 - K} \sum_{A \cap B = CI_i} m_{\text{PoA}}(A) \cdot m_{\text{CF}}(B)
\end{equation}

where \( m_{\text{PoA}}(A) \) is the belief assigned to CI \( A \) based on its PoA,  \( m_{\text{CF}}(B) \) is the belief assigned based on the CF of \( B \) and \( K \) is the conflict coefficient, representing conflicting evidence from PoA and CF:

\begin{equation}
    K = \sum_{A \cap B = \emptyset} m_{\text{PoA}}(A) \cdot m_{\text{CF}}(B)
\end{equation}

In the scenario~\ref{sec:motivating}, the CMM continuously monitors PoA and CF for the context attributes \( CA_1 \)  Presence of Road Work Sign, \( CA_2 \) Speed Limit and \( CA_4 \) Traffic Congestion Level. The CEE calculates the PoA of the \( CI_{RW} \) by considering both direct and inferred queries as shown in equation~\ref{eq:example}.

\begin{equation}
    Q_{S_1}^{\text{Total}} = Q_{S_1} + \sum_{i=1}^{n} Q_{CA_i}
    \label{eq:example}
\end{equation}

The PoA helps the system prioritize CIs. Simultaneously, the CF metric ensures that only fresh CAs remains in the cache. If the PoA of \( CI_{RW} \) is high but the CF is low (stale), the CMM refreshes the item by requesting updated information. If both PoA and CF are low, the item is evicted from the cache to free up space. If both metrics are high, the item remains cached without any update.

The \( K \) accounts for the conflicting evidence between CF and PoA, and is computed as shown in the equations below.

\begin{equation}
    K = m^i_{\text{CF}}(\text{Cache}) \cdot m^i_{\text{PoA}}(\text{Evict}) 
      + m^i_{\text{CF}}(\text{Evict}) \cdot m^i_{\text{PoA}}(\text{Cache})
\end{equation}

The combined belief masses for caching and eviction are computed as:

\begin{equation}
    m^i_{\text{Combined}}(\text{Cache}) = \frac{m^i_{\text{CF}}(\text{Cache}) \cdot m^i_{\text{PoA}}(\text{Cache})}{1 - K}
\end{equation}

\begin{equation}
    m^i_{\text{Combined}}(\text{Evict}) = \frac{m^i_{\text{CF}}(\text{Evict}) \cdot m^i_{\text{PoA}}(\text{Evict})}{1 - K}
\end{equation}

In Algorithm~\ref{alg:cache_management} CEE provides a ranked list of CIs and CAs to the CMM based on PoA. The CMM monitors both PoA and CF in real-time to manage the cache. This integration ensures that High-PoA, fresh CIs\& CAs remain in the cache. Stale but important items are refreshed. Low-priority or outdated items are evicted to optimize cache usage.

\begin{algorithm}[H]
\caption{Cache Management Algorithm Using PoA and CF}
\label{alg:cache_management}
\begin{algorithmic}[1]
\normalsize
\REQUIRE List of prioritized CIs \( C = \{C_1, C_2, \dots, C_n\} \), thresholds \( \theta_{\text{update}}, \theta_{\text{evict}} \), initialized cache
\ENSURE Optimized cache with relevant context items
\FOR{each context item \( C_i \) in cache}
    \STATE Compute \( \text{PoA}(C_i) \) using Eq.~\ref{eq:poa}
    \STATE Compute \( \text{CF}(C_i) \) using Eq.~\ref{eq:cf}
    \STATE Calculate the combined belief \( m_{\text{Combined}}(C_i) \) using Dempster-Shafer Theory (DST):
    \[
    m_{\text{Combined}}(C_i) = f(\text{PoA}(C_i), \text{CF}(C_i))
    \]
    where \( f \) is the combination function (e.g., weighted sum, DST rule).
    \IF{ \( m_{\text{Combined}}(C_i) < \theta_{\text{evict}} \) }
        \STATE Evict \( C_i \) from cache
        \STATE Log eviction: \( \text{LogEviction}(C_i) \)
    \ELSIF{ \( m_{\text{Combined}}(C_i) < \theta_{\text{update}} \) }
        \STATE Refresh \( C_i \) by requesting an update from the context provider
        \STATE Log update: \( \text{LogUpdate}(C_i) \)
    \ELSE
        \STATE Keep \( C_i \) in cache
        \STATE Log retention: \( \text{LogRetention}(C_i) \)
    \ENDIF
\ENDFOR
\STATE \textbf{Return} Updated cache and log
\end{algorithmic}
\end{algorithm}

The overall computational complexity of the framework is \(O(n \cdot m + k)\), where \(n\) is the number of context items, \(m\) is the number of attributes, and \(k\) is the number of cached items, ensuring efficient performance for real-time IoT applications.

\section{Implementation \& Evaluation in CoaaS}~\label{sec:evaluation}

In this section, we present the implementation details of the DCMF within the CoaaS platform and evaluate its performance through comprehensive experiments. The primary goals of our evaluation are \textbf{Assess Accuracy} to evaluate how accurately the DCMF model determines which CIs should be cached, replaced, or refreshed, ensuring high-quality context delivery. \textbf{Measure Efficiency} to assess the efficiency of the caching mechanism in terms of response time, cache hit ratio, cache miss ratio, cache expired ratio, and resource utilization. \textbf{Evaluate Scalability}to determine the ability of the DCMF to handle increasing volumes of CQs and CPs without significant performance degradation.

\subsection{Implementation Details}

The DCMF is integrated with the CoaaS platform developed as part of the bIoTope  project \href{https://biotope-project.eu/}{(https://biotope-project.eu/)}. The framework is implemented using a micro-services architecture in Java and Python. Communication between IoT simulators and the DCMF is handled via RESTful APIs, while inter-service communication within the DCMF utilizes gRPC for efficient data exchange.

The experiments were conducted on a system with \textbf{Processor} of Intel\textsuperscript{\textregistered} Core\texttrademark{} i7-6700 CPU @ 3.40 GHz (Turbo Boost up to 4.00 GHz), x86\_64 architecture with 8 logical cores (4 physical cores with hyper-threading). \textbf{Memory} of 32 GB DDR4 RAM @ 2133 MT/s, configured as four 8 GB DIMM modules. \textbf{Storage} as SSD for faster data access. \textbf{Software} in Ubuntu 20.04 LTS, Java 11, Python 3.8, MongoDB 4.4, SQLite 3.

The process begins when a CC sends a CQ request to CoaaS. The Context Query Engine (CQE) receives the request and disseminates it into sub-queries targeting different context entities (CEs), CIs, CAs, and situational functions as shown in Figure~\ref{DCMF}. The CQE first checks if the requested context is available in the cache. If not, it proceeds to fetch the required context from relevant CPs as described in Figure~\ref{fig:ecosystem} context sharing ecosystem.

The request details are logged in the Context Query Logs, capturing parameters such as PoA, QoS, CoC, QoC, SLAs, and timeliness. The CEE utilizes the CAPME~\cite{10643912} strategy to calculate PoA for CIs. The prioritized CIs are then passed to the Context Management Module (CMM), which employs DST to combine PoA and CF assessments. This process enables the CMM to make informed decisions on updating, evicting, or refreshing cache. By continuously monitoring the freshness of CAs and their associated belief scores, the CMM ensures that the cache contains high-quality, up-to-date, and pertinent information.

\subsection{Experimental Setup}

To evaluate the performance and scalability of the DCMF, we designed experiments with distinct configurations which involves CPs with query generation, query execution data handling and CRs pattern during roadwork.

\begin{itemize}


    \item \textbf{Context Providers and Query Generation} : We simulated CPs using the ~\href{https://github.com/IBA-Group-IT/IoT-data-simulator}{IoT Data Simulator}, incorporating real-world roadwork data captured via Nerian 3D depth cameras, shown in Figure~\ref{setup} as described in~\cite{Forkan2022}. The simulator ran locally to emulate various IoT data sources related to roadwork scenarios. CQs were generated using the~\href{https://github.com/ShakthiYasas/context-query-simulator}{Context Query Simulator}, reflecting realistic usage patterns using normal distribution.

    \begin{figure}
    \centering
    \includegraphics[width=9cm]{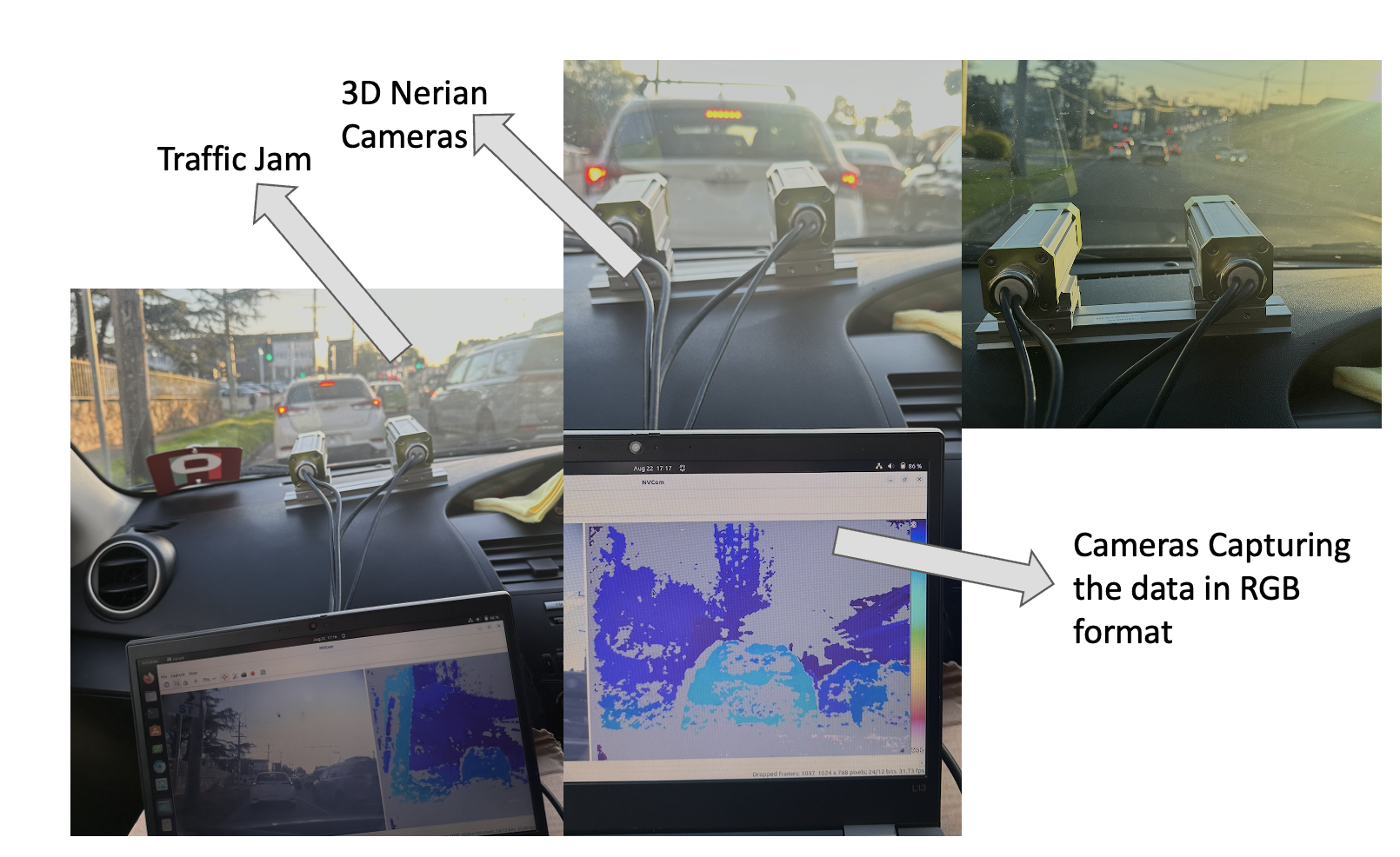}
    \caption{Road Work Data Collection setup.
    } \label{setup}
    \end{figure}

    \item \textbf{Query Execution}: A total of 70,000 CQs were processed over a 24-hour period, divided into 10 sets of 7,000 queries each to emulate real-world load conditions. The CQs were designed to simulate \textbf{low, medium, and high load conditions} using varying query arrival rates. We model these loads mathematically using Poisson distribution.

    The low, medium, and high load conditions are defined by different values of \( \lambda \) (query rate) to assess the system’s performance under varying operational scenarios \textbf{Low Load}: \( \lambda_{\text{low}} = 1 \) queries/2 second, average: 30 queries per minute; \textbf{Medium Load}: \( \lambda_{\text{medium}} = 1 \) query/second, average: 60 queries per minute \textbf{High Load}: \( \lambda_{\text{high}} = 2 \) queries/second average: 120 queries per minute. MongoDB was used for maintaining logs, statistics, and the context repository cache. SQLite was utilized for managing structured data repositories such as the context services registry, CP list, and SLAs.    
        
    \item \textbf{Dataset: Context Request Patterns During Roadwork} : The dataset used combines real-world roadwork data and publicly available online datasets for statistical analysis and training. The roadwork data, collected using Nerian 3D depth cameras in various locations, provided key information such as the location of roadwork, roadwork signs, traffic conditions, and severity levels. Supplementary data on speed zones, traffic lights, road crashes, and planned disruptions were sourced from platforms like \textit{data.world}\href{https://data.world/datasets/traffic}{data.world} and \href{https://www.land.vic.gov.au/maps-and-spatial/spatial-data/vicmap-catalogue/vicmap-transport}{Vicmap Transport}.
        
\end{itemize}

The collected dataset consists of 268,600 data points, with 75\% collected using cameras in real world setting and rest being synthetically generated using statistical distributions to enhance diversity and simulate various traffic scenarios. The dataset representing the volume of context requests during the 24-hour roadwork simulation follows a normal distribution, reflecting the temporal dynamics typical of real-world scenarios. During peak hours (6-11 AM and 3-6 PM), the mean (\(\mu\)) request rate was approximately 5,250 requests during each half-hour time slot , with a standard deviation (\(\sigma\)) of around 500 requests, indicating higher demand for real-time context during these periods. In contrast, during off-peak hours (11 AM-3 PM and 7 PM-6 AM), the mean request rate dropped to approximately 1,750 requests during each half-hour time slot, with a standard deviation (\(\sigma\)) of 300 requests, reflecting reduced demand during midday and night-time hours.

Overall, the dataset reflects a mean (\(\mu\)) of approximately 3.916, with a total of 70,000 requests processed throughout the day. The standard deviation (\(\sigma^2\)) is approximately 1.454, capturing the variability in request patterns over time. This variability is influenced by factors such as traffic conditions, time of day, and the nature of roadwork events. The observed peak request patterns align with typical commuting periods, while the fluctuations during off-peak hours demonstrate the impact of lower traffic volumes and reduced activity.

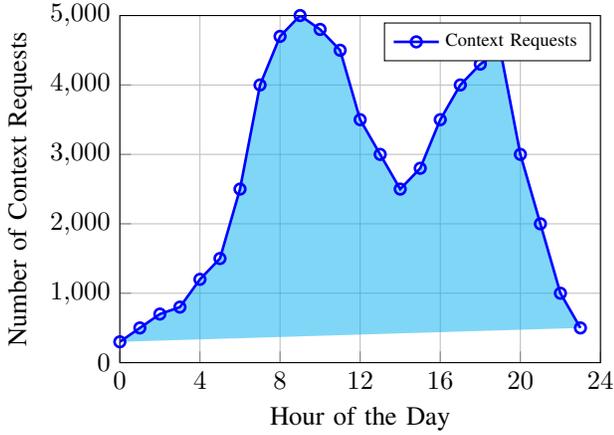
\begin{figure}
    \centering
    \begin{tikzpicture}
        \begin{axis}[
            width=0.9\columnwidth,
            height=0.7\columnwidth,
            xlabel={Hour of the Day},
            ylabel={Number of Context Requests},
            xmin=0, xmax=24,
            ymin=0, ymax=5000,
            xtick={0, 4, 8, 12, 16, 20, 24},
            ytick={0, 1000, 2000, 3000, 4000, 5000},
            legend style={at={(0.98,0.98)}, anchor=north east, font=\scriptsize},
            grid=major,
            tick align=inside,
            tick pos=left
        ]
        
        \addplot+[
            mark=o, 
            color=blue, 
            line width=1pt, 
            fill=cyan, 
            fill opacity=0.5
        ] coordinates {
            (0, 300) (1, 500) (2, 700) (3, 800) (4, 1200) (5, 1500)
            (6, 2500) (7, 4000) (8, 4700) (9, 5000) (10, 4800) (11, 4500)
            (12, 3500) (13, 3000) (14, 2500) (15, 2800) (16, 3500) (17, 4000)
            (18, 4300) (19, 4500) (20, 3000) (21, 2000) (22, 1000) (23, 500)
        };
        \addlegendentry{Context Requests}
        \end{axis}
    \end{tikzpicture}
    \caption{Context Requests Pattern Over 24 Hours}
    \label{requestPattern}
\end{figure}

Overall, Figure~\ref{requestPattern} depicts 70,000 queries, demonstrating the ability of CoaaS platform to handle varying loads and process a high volume during critical periods of roadwork. We used Apache JMeter to simulate the context-query load, setting the request rate for CQs at one per second, following a Gaussian distribution. Our setup included 30 context services, simulating approximately 10,000 users generating CQs. 

\lstset{
    language=SQL,
    basicstyle=\small\ttfamily,
    keywordstyle=\color{cyan},
    commentstyle=\color{gray},
    stringstyle=\color{red},
    identifierstyle=\color{black},
    numbers=none,
    breaklines=true,
    breakatwhitespace=true,
    showstringspaces=false,
    frame=single
}

For instance, in a road work scenario, a context-aware navigation application (CC) sends complex CDQL~\cite{s19061478} queries to the CMP requesting alternative routes as shown below. This query demonstrates how context consumers require multiple context attributes (CAs) such as location coordinates, road work status, traffic conditions, and route availability. These CAs, collected from various CPs, are often required by multiple CCs simultaneously. 
\begin{lstlisting}
PREFIX mv: http://schema.mobivoc.org, schema: http://schema.org
SELECT (alternativeRoutes.*)
DEFINE ENTITY currentLocation IS schema:Place WHERE currentLocation.address = "-37.823411, 145.042285"
ENTITY alternativeRoutes IS mv:Route WHERE (
    distance(alternativeRoutes.startPoint, currentLocation, "driving") < 500m AND
    NOT(intersects(alternativeRoutes, mv:RoadWork)) AND
    isAvailable(alternativeRoutes.availability, {start_time: now(), end_time: "2024-11-08T04:03:13"})
)
\end{lstlisting}

\subsection{Evaluation Plan and Test Scenarios}

To assess the performance of the DCMF, we developed a structured evaluation plan, including specific test scenarios to validate scalability, efficiency, and adaptability to varying operational conditions and requirements. 

\begin{itemize}
    \item \textbf{Evaluation Metrics} : The evaluation focuses on KPI described in Table~\ref{tab:performance_metrics}.

        
      \begin{table}
    \centering
    \caption{Performance Metrics and Definitions}
    \label{tab:performance_metrics}
    \begin{tabular}{|p{3cm}|p{4cm}|}
    \hline
    \textbf{Metric} & \textbf{Definition} \\ \hline
    \textbf{Cache Hit Ratio (CHR)}  & 
    
    \[
    CHR = \frac{Q_{\text{cache}}}{Q_{\text{total}}} \times 100
    \] \\ \hline
    \textbf{Cache Miss Ratio (CMR)} & 

    \[
    CMR = \frac{Q_{\text{miss}}}{Q_{\text{total}}} \times 100
    \] \\ \hline
    \textbf{Cache Expired Ratio (CER) Ratio of expired items accessed before refresh} & 
    \[
    CER = \frac{Q_{\text{expired}}}{Q_{\text{total}}} \times 100
    \] \\ \hline
    \textbf{Response Time (in milliseconds) (\(T_r\))} & 
    \[
    T_r = \frac{\sum_{i=1}^{n} t_i}{n}
    \] \\ \hline
    \textbf{Throughput per second} & 
    \[
    \text{Throughput} = \frac{Q_{\text{total}}}{T_{\text{total}}}
    \] \\ \hline
    \end{tabular}
\end{table}

    \item \textbf{Test Scenarios} : We evaluated the DCMF under four scenarios to assess its handling of CF, PoA, and scalability.

            \begin{enumerate} \item \textbf{Scenario 1: High CF Requirement}

            Simulates rapidly changing traffic conditions where CF is critical. Tests DCMF's ability to maintain low cache expired ratios and ensure prompt updates for frequently changing CAs (e.g., traffic speed, congestion levels).
            
            \item \textbf{Scenario 2: High PoA Requirement}

            Focuses on frequently accessed CIs during roadwork events. Evaluates how DCMF prioritizes high-demand CIs to achieve high cache hit ratios, reduce response times, and minimize CP load.

            \item \textbf{Scenario 3: Balanced Scenario}

            Reflects typical conditions where CF and PoA are equally important. Assesses DCMF's capability to balance freshness and high-PoA requirements, optimizing cache hit ratios and context validity.
            
            \item \textbf{Scenario 4: Scalability Test}
            
            Gradually increases CPs and CQ volume to test scalability. Evaluates DCMF's ability to maintain performance under high load by analyzing response time, throughput, cache hit ratio, and resource utilization.
            \end{enumerate}


    \item \textbf{Model Validation} : We compared the belief and utility scores generated by the DCMF's CEE and CMM against the actual caching outcomes observed during the experiments. This involved analyzing whether CIs with higher utility and belief scores corresponded to higher cache hit ratios and whether the framework correctly identified CIs that needed to be updated, refreshed, or evicted.

    \item \textbf{Fine-Tuning Model Parameters} : Based on the validation results, we adjusted key parameters within the framework to optimize performance.

    \begin{itemize}
        \item \textbf{MAUT Weightings}: We fine-tuned the weights assigned to each attribute in the MAUT calculation. Attributes such as PoA, QoS, CoC, QoC, SLAs, Context Provider Type, and Timeliness were adjusted to reflect their relative importance under different scenarios. For example, in Scenario 1, the weight for Timeliness was increased to emphasize CF.
    
        \item \textbf{Dempster-Shafer Thresholds}: We adjusted the thresholds used in the Dempster-Shafer Theory within the CMM to determine when to update, refresh, or evict context items. By fine-tuning these thresholds, we improved the framework's responsiveness to changes in CF and PoA, enhancing overall caching efficiency.   The thresholds were determined using historical query patterns and access logs:
  \[
  \theta_{\text{update}} = \mu_{\text{CF}} - \kappa \cdot \sigma_{\text{CF}}
  \]
  \[
  \theta_{\text{evict}} = \mu_{\text{CF}} - 2 \cdot \kappa \cdot \sigma_{\text{CF}}
  \]
  Where:
  - \( \mu_{\text{CF}} \): Mean CF score for all cached items.
  - \( \sigma_{\text{CF}} \): Standard deviation of CF scores.
  - \( \kappa \): A sensitivity parameter to control the frequency of updates/evictions.
    \end{itemize}

\end{itemize}

\subsection{Comparison with State-of-the-Art Techniques}

To evaluate the proposed DCMF, we compared its performance against state-of-the-art caching algorithms tailored for context-aware systems: \textbf{m-CAC}, \textbf{m-Greedy}, and \textbf{m-Myopic}~\cite{Muller2017}. These strategies serve as benchmarks for Random and Least Recently Used (LRU) caching policies, respectively.

The m-CAC algorithm employs a multi-armed bandit (MAB) approach to balance exploration and exploitation in caching decisions, maximizing cache hit rates through context and user access patterns. While originally applied to content delivery networks~\cite{Muller2017}, we adapted its utility function to align with the dynamic nature of IoT data in roadwork scenarios.

\paragraph{Adaptation to IoT Roadwork Scenario}
The original m-CAC utility function prioritizes popularity (\(P(c_i)\)) and relevance (\(R(c_i)\)) as:

\begin{equation}
U_{\text{original}}(c_i) = \alpha \cdot P(c_i) + (1 - \alpha) \cdot R(c_i),
\label{eq:original_utility}
\end{equation}

where \( \alpha \) balances the two factors. In our roadwork adaptation, For roadwork scenario in IoT environments, we adapted the utility function to incorporate factors relevant to dynamic context. The adapted utility \( U_{\text{adapted}}(c_i) \) is defined as:

\begin{IEEEeqnarray}{rCl}
U_{\text{adapted}}(c_i) & = & w_1 \cdot \text{PoA}(c_i) + w_2 \cdot \text{CF}(c_i) \nonumber\\
& & {} +\ w_3 \cdot \text{QoS}(c_i) + w_4 \cdot \text{Timeliness}(c_i)
\label{eq:adapted_utility}
\end{IEEEeqnarray}

where \( \text{PoA}(c_i) \) is the \textbf{PoA} for context item \( c_i \), \( \text{CF}(c_i) \) is the \textbf{CF} of \( c_i \),  \( \text{QoS}(c_i) \) is the \textbf{QoS} associated with \( c_i \), \( \text{Timeliness}(c_i) \) reflects how \textbf{timely} the context item \( c_i \) is and \( w_1, w_2, w_3, w_4 \) are weighting factors such that \( \sum_{i=1}^{4} w_i = 1 \).

\paragraph{Assumptions in m-CAC}

The m-CAC algorithm operates under several assumptions:

\begin{itemize}
    \item \textbf{Availability and Accuracy of Context}: Assumes comprehensive and accurate context information is readily available for each request.
    \item \textbf{Stationary Context-Popularity Relationship}: Assumes that the relationship between context and item popularity remains relatively stable over time.
    \item \textbf{Fixed Cache Size (\( m \))}: Assumes a known and fixed cache capacity that can hold exactly \( m \) items.
    \item \textbf{Uniform Item Size}: Assumes all cached items are of the same size or that size differences are negligible.
\end{itemize}

This adaptation ensures that the caching decisions are contextually relevant to the dynamic nature of IoT data in smart city applications. By incorporating these factors, we maintain the original algorithmic framework of m-CAC but tailor the utility function to better suit the requirements of our scenario.

For a comprehensive evaluation, we also included the m-Greedy and m-Myopic algorithms in our comparisons. The m-Greedy algorithm is a caching strategy that prioritizes items based on immediate rewards, focusing on frequently accessed content. The m-Myopic algorithm is a context-aware version of the Myopic policy, similar to the Least Recently Used (LRU) strategy. It makes caching decisions based on the most recent access information, caching items.

\section{Results \& Discussion}~\label{sec:result}

The evaluation of the DCMF was conducted across four scenarios designed to highlight its dual focus on PoA and CF. Each scenario reflects distinct operational conditions in IoT environments, with Gaussian distributions used to simulate realistic IoT data patterns and access behaviors.

    
    
    

\subsection{Scenario 1: High CF Requirement}

In Scenario 1, we simulate a roadwork use case characterized by rapidly changing traffic conditions during peak construction periods, where the likelihood of hazards or accidents is elevated. This scenario demands high CF to ensure that CCs receive the most up-to-date information for timely decision-making. Maintaining high CF in dynamic IoT environments ensures timely updates to users, reducing the risks associated with outdated information. This is particularly critical during periods of rapid change, where stale context could lead to suboptimal decision-making and inefficiencies.


To evaluate the DCMF's performance under these conditions, we prioritized CF in the utility function, assigning higher weights to the CF (CF) attribute. The framework's Context Management Module (CMM) actively monitored the freshness of CIs, ensuring that stale data was promptly updated or evicted from the cache.

\begin{figure}
    \centering
    \begin{tikzpicture}
        \begin{axis}[
             width=0.9\columnwidth,
            height=0.7\columnwidth,
            xlabel={Cache Size (MB)},
            ylabel={Cumulative Cache Hits ($\times 10^4$)},
            xmin=50, xmax=1000, 
            ymin=50000, ymax=250000, 
            ytick={50000,100000,150000,200000,250000},
            yticklabels={5, 10, 15, 20, 25},
            legend style={at={(0.4,0.98)}, anchor=north east},
            grid=major,
            tick align=inside,
            tick pos=left,
        ]
        
        \addplot[mark=o, color=blue] coordinates {
            (50, 60000)
            (250, 120000)
            (500, 180000)
            (750, 210000)
            (1000, 230000)
        };
        \addlegendentry{DCMF}
        
        \addplot[mark=triangle, color=green] coordinates {
            (50, 55000)
            (250, 110000)
            (500, 160000)
            (750, 195000)
            (1000, 215000)
        };
        \addlegendentry{m-CAC}
        
        \addplot[mark=diamond, color=purple] coordinates {
            (50, 50000)
            (250, 105000)
            (500, 155000)
            (750, 185000)
            (1000, 205000)
        };
        \addlegendentry{m-Greedy}
        
        \addplot[mark=square, color=red] coordinates {
            (50, 45000)
            (250, 95000)
            (500, 140000)
            (750, 170000)
            (1000, 190000)
        };
        \addlegendentry{m-Myopic}
        
        \end{axis}
    \end{tikzpicture}
    \caption{Cumulative Cache Hits vs. Cache Size for Different Caching Techniques}
        \label{fig1}
\end{figure}
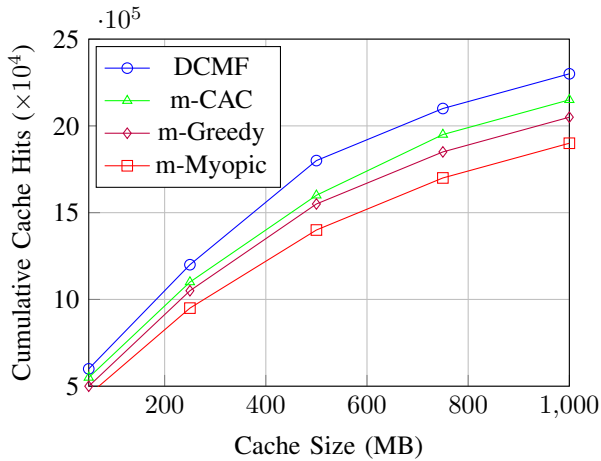

\begin{figure}
    \centering
    \begin{tikzpicture}
        \begin{axis}[
            width=0.45\textwidth, 
            height=0.35\textwidth, 
            xlabel={Cache Size (MB)},
            ylabel={Cache Expired Ratio},
            xmin=0, xmax=1000,
            ymin=0, ymax=0.5,
            xtick={0,200,400,600,800,1000},
            ytick={0.0, 0.1, 0.2, 0.3, 0.4, 0.5},
            grid=major,
           legend style={at={(0.8,0.97)}, anchor=north, font=\small}, 
            thick,
            tick label style={font=\small},
            xlabel ,
            ylabel ,
            scaled y ticks = false
        ]
        
        \addplot[color=blue, mark=o] 
            coordinates {(0,0.12) (200,0.11) (400,0.10) (600,0.09) (800,0.09) (1000,0.09)};
        \addlegendentry{DCMF}
        
        \addplot[color=green, mark=square*, mark size=2pt] 
            coordinates {(0,0.30) (200,0.22) (400,0.18) (600,0.15) (800,0.13) (1000,0.12)};
        \addlegendentry{m-CAC}
        
        \addplot[color=orange, mark=triangle*, mark size=2pt] 
            coordinates {(0,0.35) (200,0.28) (400,0.25) (600,0.22) (800,0.18) (1000,0.18)};
        \addlegendentry{m-Greedy}
        
        \addplot[color=red, mark=x, mark size=2pt] 
            coordinates {(0,0.40) (200,0.30) (400,0.27) (600,0.25) (800,0.23) (1000,0.22)};
        \addlegendentry{m-Myopic}
        
        \end{axis}
    \end{tikzpicture}
    \caption{Cache Expired Ratio for Different Cache Sizes (MB)}
    \label{fig2}
\end{figure}
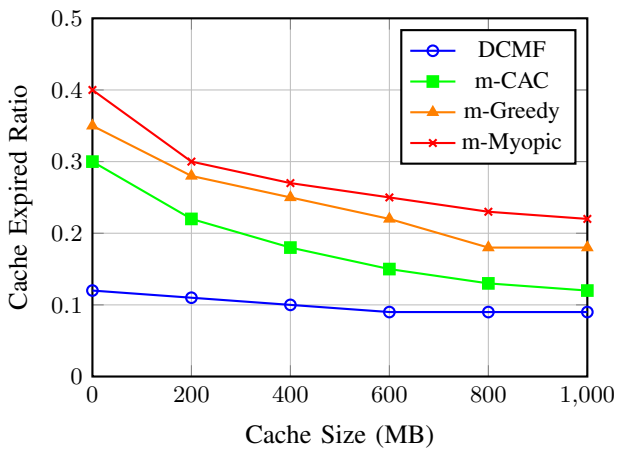

The results, shown in Figure~\ref{fig1}, indicate that DCMF achieved a cache hit ratio of approximately 80\%, significantly higher than the 65\% achieved by m-CAC and even lower percentages by m-Greedy and m-Myopic. The cache expired ratio for DCMF was as low as 5\%, as shown in Figure~\ref{fig2}, indicating effective freshness maintenance, whereas the comparison algorithms exhibited expired ratios ranging from 18\% to 25\%. The average response time for DCMF was reduced to 120 ms due to the efficient retrieval of fresh context from the cache, compared to 190 ms for m-CAC.

As seen in Figure~\ref{fig1}, the performance gap between DCMF and other techniques widens with increasing cache size. This is due to DCMF's ability to prioritize frequently updated and accessed context items, ensuring that the cache holds high-value items. In contrast, m-CAC and m-Greedy exhibit lower adaptability to dynamic updates, resulting in higher cache misses and expired items.

Compared to m-CAC, DCMF showed a 23.1\% improvement in cache hit ratio and a 36.8\% reduction in average response time. The variance in response times for DCMF (\( \sigma = 15 \, \mathrm{ms} \)) was also significantly lower than for m-CAC (\( \sigma = 32 \, \mathrm{ms} \)).

One limitation observed in this scenario was a slight increase in memory usage compared to baseline techniques due to the frequent refresh of context items. However, this trade-off is justified by the significant improvement in CF and reduced response times.


The graph illustrates that DCMF maintains a higher proportion of fresh context in the cache, leading to improved performance in environments where context changes rapidly. This demonstrates the framework's capability to handle scenarios requiring immediate and accurate context updates, which is critical in mitigating risks associated with roadwork hazards and accidents.

\subsection{Scenario 2: High PoA Requirement}

In Scenario 2, we analyze situations where specific CIs generate a surge in information requests due to their significant impact on traffic flow. Examples include major construction projects, lane closures during peak hours, or emergency repairs following accidents. These events lead to a high PoA for related CIs, as numerous CCs frequently query for updates on these disruptions.

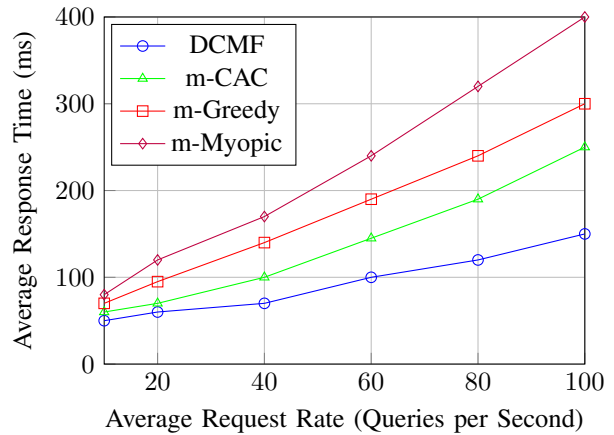
\begin{figure}
    \centering
    \begin{tikzpicture}
        \begin{axis}[
            width=0.9\columnwidth,
            height=0.7\columnwidth,
            xlabel={Average Request Rate (Queries per Second)},
            ylabel={Average Response Time (ms)},
            xmin=10, xmax=100,
            ymin=0, ymax=400,
            legend style={at={(0.4,0.98)}, anchor=north east},
            grid=major,
            tick align=inside,
            tick pos=left,
        ]
        
        \addplot[mark=o, color=blue] coordinates {
            (10, 50) (20, 60) (40, 70) (60, 100) (80, 120) (100, 150)
        };
        \addlegendentry{DCMF}
        
        \addplot[mark=triangle, color=green] coordinates {
            (10, 60) (20, 70) (40, 100) (60, 145) (80, 190) (100, 250)
        };
        \addlegendentry{m-CAC}
        
        \addplot[mark=square, color=red] coordinates {
            (10, 70) (20, 95) (40, 140) (60, 190) (80, 240) (100, 300)
        };
        \addlegendentry{m-Greedy}
        
        \addplot[mark=diamond, color=purple] coordinates {
            (10, 80) (20, 120) (40, 170) (60, 240) (80, 320) (100, 400)
        };
        \addlegendentry{m-Myopic}
        
        \end{axis}
    \end{tikzpicture}
    \caption{Average Response Time vs. Request Rate for Different Caching Techniques}
    \label{fig:response_vs_rate}
\end{figure}

To simulate this scenario, we modeled frequent access to critical context information using a Gaussian distribution. The distribution was centered on the peak demand times (i.e. 8 to 10 AM and 5 to 7 PM), representing increased interest during periods of high activity. For instance, during the simulated peak periods, the query rate averaged 120 queries per second, with a standard deviation of 15 queries, reflecting the increased demand. In contrast, during non-peak periods (for example, 11AM - 4PM), the average query rate dropped to 50 queries per second with a standard deviation of 10 queries.

The DCMF's CAPME dynamically calculates PoA by analyzing query patterns in real time. Using these query rates, the utility scores in the MAUT framework were adjusted to prioritize high-PoA CIs. For instance, CIs with query counts exceeding 1,000 during peak hours were assigned a significantly higher utility score compared to items accessed less frequently. This adjustment ensured that frequently queried items remained cached.

Figure~\ref{fig:response_vs_rate} shows that the DCMF achieves the lowest average response time across varying request rates, highlighting its ability to efficiently cache high-demand CIs. At a maximum request rate of 100 queries per second, DCMF maintains an average response time of 150 ms, outperforming m-CAC (250 ms), m-Greedy (300 ms) and m-Myopic (400 ms). This performance is crucial for systems that require real-time response.

\begin{figure}
    \centering
    \begin{tikzpicture}
        \begin{axis}[
            width=0.9\columnwidth,
            height=0.7\columnwidth,
            xlabel={Cache Size (MB)},
            ylabel={Average Latency (ms)},
            xmin=50, xmax=1000,
            ymin=10, ymax=35,
            xtick={50, 250, 500, 750, 1000},
            ytick={10, 15, 20, 25, 30, 35},
            legend style={at={(0.98,0.98)}, anchor=north east},
            grid=major,
            tick align=inside,
            tick pos=left,
        ]
        
        \addplot[mark=o, color=blue] coordinates {
            (50, 22) (250, 20) (500, 18) (750, 15) (1000, 14)
        };
        \addlegendentry{DCMF}
        
        \addplot[mark=triangle, color=green] coordinates {
            (50, 27) (250, 24) (500, 21) (750, 18) (1000, 16)
        };
        \addlegendentry{m-CAC}
        
        \addplot[mark=triangle*, color=orange] coordinates {
            (50, 30) (250, 28) (500, 26) (750, 24) (1000, 22)
        };
        \addlegendentry{m-Greedy}
        
        \addplot[mark=x, color=red] coordinates {
            (50, 32) (250, 30) (500, 27) (750, 25) (1000, 23)
        };
        \addlegendentry{m-Myopic}
        
        \end{axis}
    \end{tikzpicture}
    \caption{Average Latency vs. Cache Size for Different Caching Techniques}
    \label{fig:latency_vs_cache}
\end{figure}
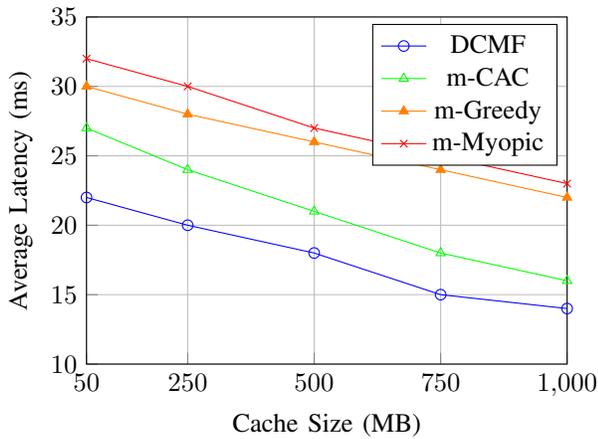

Figure~\ref{fig:latency_vs_cache} further illustrates the effectiveness of DCMF by showing lower average latency across varying cache sizes. With a 1000 MB cache, DCMF maintains a latency of 14 ms, compared to m-CAC (16 ms), m-Greedy (22 ms), and m-Myopic (23 ms). This reduction in latency ensures that frequently accessed context items remain readily available, enhancing system performance.

By emphasizing PoA in the utility function, DCMF optimizes cache usage, achieving an 85\% cache hit ratio, significantly higher than m-CAC (68\%), m-Greedy (62\%), and m-Myopic (58\%). The higher cache hit ratio minimizes cache misses, reduces dependency on CPs, and ensures real-time responses to critical CQs. These metrics underscore DCMF’s ability to handle high-demand scenarios efficiently, maintaining system responsiveness and improving user experience.

Through effective prioritization of high-PoA CIs, the DCMF demonstrates superior performance in scenarios where rapid and frequent access to critical information is essential. This capability ensures timely decision making for IoT applications in dynamic IoT environments.

\subsection{Scenario 3: Balanced Scenario of CF \& PoA}

In Scenario 3, we consider typical operating conditions in a dynamic IoT environment where both CF and PoA are equally important. This scenario reflects realistic conditions where CCs require timely and frequently accessed information to make informed decisions, such as updates on traffic conditions, alternate routes due to ongoing disruptions, or scheduled maintenance activities.

To emulate this balanced scenario, we configured the DCMF to assign equal weights (\(w_{\text{CF}} = w_{\text{PoA}} = 0.5\)) to CF and PoA in the utility function within the MAUT framework. A Gaussian distribution was used to simulate moderate variability in access patterns, representing everyday fluctuations in urban networks. The simulated query arrival rate ranged between 50 and 100 queries per second, with an average query rate of 75 queries per second during normal conditions.

The DCMF’s dual-engine approach dynamically adjusts caching decisions to ensure that cached context remains both fresh and frequently requested. CAPME monitors access frequencies, while the CMM evaluates CF to ensure that cached information is up-to-date as shown in Table~\ref{tab:balanced_metrics}.

\begin{table}
    \centering
    \caption{Performance Metrics in Scenario 3}
    \label{tab:balanced_metrics}
    \begin{tabular}{|p{2cm}|p{1cm}|p{1cm}|}
    \hline
    \textbf{Metric} & \textbf{DCMF} & \textbf{m-CAC} \\ \hline
    Cache Hit Ratio (CHR) & 82\% & 68\% \\ \hline
    Cache Expired Ratio (CER) & 6\% & 15\% \\ \hline
    Average Response Time & 120 ms & 190 ms \\ \hline
    Resource Utilization & 72\% & 81\% \\ \hline
    \end{tabular}
\end{table}

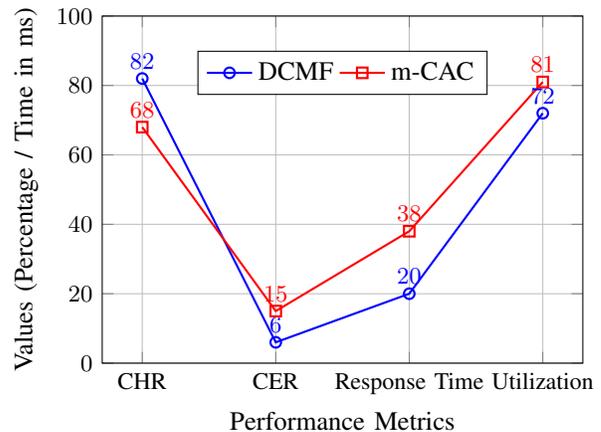
\begin{figure}
    \centering
    \begin{tikzpicture}
        \begin{axis}[
            width=0.9\columnwidth,
            height=0.7\columnwidth,
            xlabel={Performance Metrics},
            ylabel={Values (Percentage / Time in ms)},
            symbolic x coords={CHR, CER, Response Time, Utilization},
            xtick=data,
            ymin=0, ymax=100,
            grid=major,
            legend style={at={(0.5,0.9)}, anchor=north, legend columns=2},
            nodes near coords,
            every node near coord/.append style={font=\small},
            tick label style={font=\small},
        ]
        \addplot[
            color=blue, 
            mark=o, 
            thick
        ] coordinates {
            (CHR, 82) (CER, 6) (Response Time, 20) (Utilization, 72)
        };
        \addlegendentry{DCMF}

        \addplot[
            color=red, 
            mark=square, 
            thick
        ] coordinates {
            (CHR, 68) (CER, 15) (Response Time, 38) (Utilization, 81)
        };
        \addlegendentry{m-CAC}
        \end{axis}
    \end{tikzpicture}
    \caption{Performance Comparison for Scenario 3: Line Graph}
    \label{fig:scenario3_linegraph}
\end{figure}

Figure~\ref{fig:scenario3_linegraph} visualizes the key performance metrics. As shown, DCMF achieves higher cache hit ratios and lower expired ratios compared to m-CAC, demonstrating its effectiveness in maintaining a balance between CF and PoA. Additionally, DCMF outperforms m-CAC in terms of response time and resource utilization, indicating better scalability and efficiency.

\subsection{Scenario 4: Scalability Test}

In Scenario 4, we evaluate the scalability of the DCMF by analyzing its performance under increasing numbers of CPs and CQs. This scenario simulates high-load conditions typical in large-scale IoT applications, where numerous CPs continuously share context , and multiple CCs query for relevant information simultaneously.

To simulate this scenario, the number of CPs was progressively increased from 30 to 100, and the CQ rate was scaled accordingly to emulate a surge in both data availability and user demand. Gaussian distributions were used to model the arrival patterns of the CQ, reflecting realistic variations over time.

\begin{figure}
    \centering
    \begin{tikzpicture}
        \begin{axis}[
            width=0.9\columnwidth,
            height=0.6\columnwidth,
            xlabel={Caching Techniques},
            ylabel={Throughput (Requests per Minute)},
            symbolic x coords={DCMF, m-CAC, m-Myopic, m-Greedy},
            xtick=data,
            ymin=30, ymax=60,
            bar width=8pt,
            ymajorgrids,
            legend style={at={(0.5,1.2)}, anchor=north, legend columns=3},
            nodes near coords,
            every node near coord/.append style={font=\small},
        ]
        \addplot [
            color=red, 
            mark=square, 
            thick
        ] coordinates {(DCMF, 53.8) (m-CAC, 45.76) (m-Myopic, 41.11) (m-Greedy, 36.82)};
        \addlegendentry{Low Load}
        
        \addplot [
            color=blue, 
            mark=o, 
            thick
        ] coordinates {(DCMF, 50.5) (m-CAC, 43.1) (m-Myopic, 39.2) (m-Greedy, 34.6)};
        \addlegendentry{Medium Load}
        
        \addplot [
            color=orange, 
            mark=triangle, 
            thick
        ] coordinates {(DCMF, 47.8) (m-CAC, 41.8) (m-Myopic, 37.5) (m-Greedy, 33.0)};
        \addlegendentry{High Load}
        \end{axis}
    \end{tikzpicture}
    \caption{Throughput per Minute Comparison of Caching Techniques under Different Load Conditions}
    \label{fig:throughput_comparison}
\end{figure}
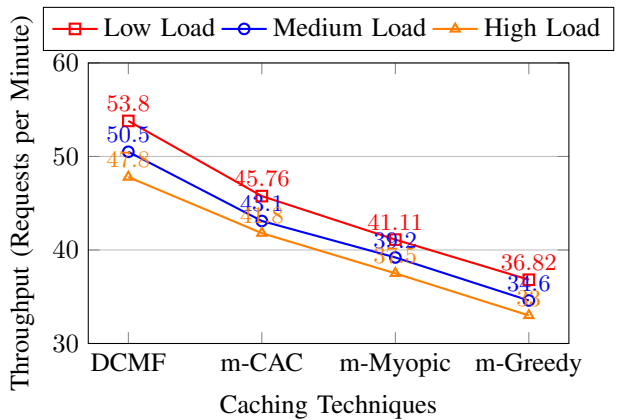

Figure~\ref{fig:throughput_comparison} compares the throughput of various caching techniques under low, medium, and high load conditions. DCMF consistently achieves the highest throughput across all load levels, with 53.8 requests per minute under low load, demonstrating its superior capability to handle increasing demands efficiently.



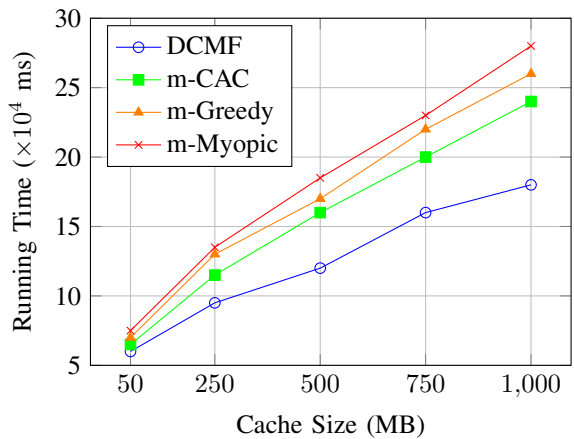
\begin{figure}
    \centering
    \begin{tikzpicture}
        \begin{axis}[
            width=0.9\columnwidth,
            height=0.7\columnwidth,
            xlabel={Cache Size (MB)},
            ylabel={Running Time ($\times 10^4$ ms)},
            ymin=50000, ymax=300000,
            ytick={50000,100000,150000,200000,250000,300000},
            yticklabels={5, 10, 15, 20, 25, 30},
            xtick={50, 250, 500, 750, 1000},
            legend style={at={(0.42,0.98)}, anchor=north east, draw=black},
            legend cell align={left},
            grid=major,
            scaled y ticks=false,
        ]
        
        \addplot[color=blue, mark=o] 
            coordinates {(50,60000) (250,95000) (500,120000) (750,160000) (1000,180000)};
        \addlegendentry{DCMF}
        
        \addplot[color=green, mark=square*] 
            coordinates {(50,65000) (250,115000) (500,160000) (750,200000) (1000,240000)};
        \addlegendentry{m-CAC}
        
        \addplot[color=orange, mark=triangle*] 
            coordinates {(50,70000) (250,130000) (500,170000) (750,220000) (1000,260000)};
        \addlegendentry{m-Greedy}
        
        \addplot[color=red, mark=x] 
            coordinates {(50,75000) (250,135000) (500,185000) (750,230000) (1000,280000)};
        \addlegendentry{m-Myopic}
        
        \end{axis}
    \end{tikzpicture}
    \caption{Running Time for Different Cache Sizes (MB)}
    \label{fig:running_time}
\end{figure}

Furthermore, as shown in Figure~\ref{fig:running_time}, DCMF sustained a low running time across varying cache sizes, demonstrating superior efficiency in managing cache operations. For instance, at 1000 MB cache size, the running time for DCMF remained under 180,000 ms, compared to m-Myopic, which reached 280,000 ms. This indicates that DCMF effectively handles increased cache sizes and query volumes without significant performance degradation.


The results of Figures~\ref{fig:throughput_comparison} and \ref{fig:running_time} highlight the scalability advantages of DCMF. The framework’s adaptive caching strategies, efficient data management, and parallel processing capabilities ensure consistent performance even under high-load conditions. 


\section{Conclusion}

In this paper, we presented the DCMF, a novel solution for efficient and adaptive caching in IoT environments. The DCMF integrates advanced decision-making theories, namely MAUT and DST, to evaluate and manage context based on multiple attributes such as PoA, CF, QoS, CoC, QoC and Timeliness. By addressing the unique challenges of dynamic contexts and resource constraints in IoT applications, the DCMF enhances the delivery of high-quality context information to CCs.

Our extensive evaluation, structured around four test scenarios—high CF requirement, high PoA requirement, balanced conditions, and scalability testing—demonstrated that the DCMF significantly outperforms state-of-the-art caching algorithms like m-CAC, m-Greedy, and m-Myopic. The results showed substantial improvements in cache hit ratios, reduced cache expired ratios, lower average response times, and efficient resource utilization. The framework effectively adapts to varying conditions, maintaining performance even under increased load, which confirms its suitability for large-scale IoT deployments. Future work includes exploring machine learning integration for predictive context management, enhancing security and privacy features, and optimizing energy efficiency for resource-constrained devices.

\section*{Acknowledgment}
Support for this publication from the Australian Research Council (ARC) Discovery Project Grant DP200102299 is thankfully acknowledged.

\end{document}